\documentclass[aps,twocolumn,showpacs,byrevtex]{revtex4-1}
\usepackage{times}
\usepackage{amsmath}
\usepackage{amsfonts}
\usepackage{amssymb}
\usepackage{amsthm}
\usepackage{bm}
\usepackage{latexsym}
\usepackage{graphicx}
\usepackage{color}
\usepackage{url}
\usepackage{subfigure}

\newcommand{\pip}{\pi^{+}}
\newcommand{\pim}{\pi^{-}}
\newcommand{\piz}{\pi^{0}}

\newcommand{\ks}{K^0_S}
\newcommand{\ar}{\rightarrow}

\newcommand{\GeV}{GeV/$c^2$}
\newcommand{\MeV}{MeV/$c^2$}

\newcommand{\BR}{\mathcal{B}}
\newcommand{\dm}{~$D$~}
\newcommand{\mbc}{M_\mathrm{BC}}
\newcommand{\delE}{\Delta E}
\newcommand{\dEdx}{\ensuremath{\mathrm{d}E/\mathrm{d}x}}

\newcommand{\dpomegapi}{D^{+}\to\omega\pi^{+}}
\newcommand{\dzomegapi}{D^{0}\to\omega\piz}
\newcommand{\dpetapi}{D^{+}\to\eta\pi^{+}}
\newcommand{\dzetapi}{D^{0}\to\eta\piz}

\newcommand{\sigDpStat}{5.5\sigma}
\newcommand{\sigDzStat}{4.1\sigma}
\newcommand{\resultBrDp}{2.79\pm0.57\pm0.16}
\newcommand{\resultBrDz}{1.17\pm0.34\pm0.07}

\newcommand{\normalBrDpStatEta}{3.07\pm0.22}
\newcommand{\normalBrDzStatEta}{0.65\pm0.09}
\newcommand{\resultBrDpEta}{\normalBrDpStatEta\pm0.13}
\newcommand{\resultBrDzEta}{\normalBrDzStatEta\pm0.04}

\newcommand{\BrDpEtaPDG}{3.53\pm0.21}
\newcommand{\BrDzEtaPDG}{0.68\pm0.07}

\usepackage[pagewise]{lineno}

\begin{document}

\title{\boldmath Observation of the Singly Cabibbo-Suppressed Decay $\dpomegapi$ and Evidence for $\dzomegapi$ }

\author{
  \begin{small}
    \begin{center}
      M.~Ablikim$^{1}$, M.~N.~Achasov$^{9,f}$, X.~C.~Ai$^{1}$,
      O.~Albayrak$^{5}$, M.~Albrecht$^{4}$, D.~J.~Ambrose$^{44}$,
      A.~Amoroso$^{49A,49C}$, F.~F.~An$^{1}$, Q.~An$^{46,a}$,
      J.~Z.~Bai$^{1}$
% ...
\\
\vspace{0.2cm}
(BESIII Collaboration)\\
\vspace{0.2cm} {\it
$^{1}$ Institute of High Energy Physics, Beijing 100049, People's Republic of China\\
% ...
$^{4}$ Bochum Ruhr-University, D-44780 Bochum, Germany\\
$^{5}$ Carnegie Mellon University, Pittsburgh, Pennsylvania 15213, USA\\
% ...
$^{9}$ G.I. Budker Institute of Nuclear Physics SB RAS (BINP), Novosibirsk 630090, Russia\\
% ...
$^{44}$ University of Rochester, Rochester, New York 14627, USA\\
% ...
$^{46}$ University of Science and Technology of China, Hefei 230026, People's Republic of China\\
% ...
$^{49}$ (A)University of Turin, I-10125, Turin, Italy; (B)University of Eastern Piedmont, \\I-15121, Alessandria, Italy; (C)INFN, I-10125, Turin, Italy\\
% ...
$^{a}$ Also at State Key Laboratory of Particle Detection and Electronics, Beijing 100049, Hefei 230026, People's Republic of China\\
% ...
$^{f}$ Also at the Novosibirsk State University, Novosibirsk, 630090, Russia\\
% ...
}
\end{center}
\vspace{0.4cm}
\end{small}
} 
\affiliation{}

\date{\today}

\vspace{0.4cm}
\begin{abstract}

Based on 2.93 fb$^{-1}$ $e^+e^-$ collision data taken at center-of-mass energy of 3.773 GeV by the BESIII detector,
we report searches for the singly Cabibbo-suppressed decays $\dpomegapi$ and $\dzomegapi$.
A double tag technique is used to measure the
absolute  branching fractions $\BR(\dpomegapi)=(\resultBrDp)\times 10^{-4}$
and $\BR(\dzomegapi)=(\resultBrDz)\times 10^{-4}$, with statistical
significances of $\sigDpStat$ and $\sigDzStat$, respectively.
We also present measurements of the absolute branching fractions for
the related $\eta \pi$ decay modes.
We find $\BR(\dpetapi)=(\resultBrDpEta)\times10^{-3}$ and
$\BR(\dzetapi)=(\resultBrDzEta)\times10^{-3}$,
which are consistent with the current world averages.
The first and second uncertainties are statistical and systematic,
respectively.

\end{abstract}


\maketitle

Hadronic decays of charm mesons provide important input for beauty physics
and also open a window into the study of strong final state interactions.
For Cabibbo-suppressed charm decays, precise measurements are challenging
due to low statistics and high backgrounds.
Among them, the singly Cabibbo-suppressed (SCS) decays $D^{+,0} \to
\omega\pi^{+,0}$ have not yet been observed.
The most recent experimental search was performed by the
CLEO Collaboration in 2006~\cite{4piCLEO} with
a 281 pb$^{-1}$ data collected on the $\psi(3770)$ peak. The branching ratio upper limits were set to be $3.4\times10^{-4}$ and $2.6\times10^{-4}$ at the $90\%$
confidence level (C.L.) for $\dpomegapi$ and $\dzomegapi$, respectively~\cite{4piCLEO}. Following the diagrammatic approach,
the small decay rates may be caused by the destructive interference between the color-suppressed
quark diagrams $C_V$ and $C_P$~\cite{HYCheng}. Numerically,
if $W$-annihilation contributions are neglected, the branching fractions of
the $D\to\omega\pi$ decays should be at about $1.0\times10^{-4}$ level~\cite{HYCheng,HYChengDiscussion}.

Besides searching for $D^{+,0}\to \omega \pi^{+,0}$, we also report measurements of the branching fractions for the
decays $D^{+,0}\to \eta \pi^{+,0}$. Precise measurements of these decay rates can improve understanding of $U$-spin and
$SU(3)$-flavor symmetry breaking effects in $D$ decays, benefiting theoretical predictions of $CP$ violation in $D$
decays~\cite{Grossman:2012ry}.

We employ the ``double tag'' (DT) technique first developed by the MARK-III Collaboration~\cite{DTMarkIII1,DTMarkIII2} to perform absolute measurements of the
branching fractions.
As the peak of the $\psi(3770)$ resonance  is just above the $D\bar{D}$ threshold and below the $D\bar{D}\pi$ threshold,
for $D$ meson we are interested, only $D\bar{D}$ pair-production is allowed. We select ``single tag'' (ST)
events in which either a $D$ or $\bar{D}$ is fully reconstructed without
reference to the other meson.
We then look for the $D$ decays of interest in the remainder of each event,
namely, in DT events where both the $D$ and $\bar{D}$ are fully reconstructed.
 This strategy suppresses background and provides an absolute normalization for branching fraction measurements without the need for knowledge of the luminosity or the $e^+e^- \to D\bar{D}$ production cross section.
 The absolute
branching fractions for $D$ meson decays are calculated by the general formula
\begin{equation}\label{forAbsBR}
\BR_{\rm sig} =
   \frac{
         \sum_{\alpha}N_{\rm sig}^{\rm obs,\alpha}
        }
        {
          \sum_{\alpha}N_{\rm tag}^{\rm obs,\alpha}
          \epsilon_{\rm tag,sig}^{\alpha}
         /\epsilon_{\rm tag}^{\alpha}
        },
\end{equation}
where $\alpha$ denotes different ST modes,
$N_{\rm tag}^{\rm obs,\alpha}$ is the yield of ST events for the tag
mode $\alpha$, $N_{\rm sig}^{\rm obs,\alpha}$ is the corresponding
yield of DT events, and $\epsilon_{\rm tag}^{\alpha}$ and $\epsilon_{\rm tag,sig}^{\alpha}$ are the ST and DT efficiencies for the tag mode $\alpha$ .

BESIII is a general-purpose magnetic spectrometer
with a helium-gas-based drift chamber (MDC), a plastic
scintillator time-of-flight system (TOF), and a CsI(Tl)
electromagnetic calorimeter (EMC) enclosed in a superconducting
solenoidal magnet providing a 1.0~T field.
The solenoid is supported by an octagonal flux-return
yoke with resistive-plate counters interleaved with steel
for muon identification (MUC). The acceptance for
charged particles and photons is 93\% of 4$\pi$, and the
charged particle momentum and barrel (endcap) photon energy resolutions
at 1 GeV are 0.5\% and 2.5\% (5.0\%), respectively~\cite{bes3}.
The data used has  an integrated luminosity of 2.93\,fb$^{-1}$~\cite{lumpsipp}
and was collected with the BESIII detector at a center-of-mass energy
of 3.773 GeV.


A {\sc geant4}-based~\cite{geant4} Monte-Carlo (MC) simulation package,
which includes the geometric
description of the detector and the detector response, is used to determine
the detection efficiency and to estimate the potential peaking background.
Signal MC samples of a $D$ meson decaying only to $\omega\pi$ ($\eta \pi$)
together with
a $\bar{D}$ decaying only to the tag modes used are generated by the MC
generator {\sc kkmc}~\cite{kkmc} using {\sc evtgen}~\cite{evtgen},
with initial state radiation (ISR) effects~\cite{isr} and
final state radiation effects~\cite{photons} included.
For the background studies, MC samples of
$\psi(3770) \to D^0\bar D^0, D^+D^-$ and $\psi(3770) \to$
non-$D\bar D$ decays, ISR production of $\psi(3686)$ and $J/\psi$,
and $e^+e^- \to q\bar q$ continuum processes, are produced at $\sqrt s=3.773$\,GeV.
All known decay modes of the various $D$ and $\psi$ mesons are generated
with branching fractions taken from the Particle Data Group
(PDG)~\cite{pdg2012},
and the remaining decays are generated with {\sc lundcharm}~\cite{lundcharm}.


Charged tracks are required to be well-measured and to
satisfy criteria based on the track fit quality;
the angular range is restricted to $|\cos\theta|<0.93$, where $\theta$
is the polar angle with respect to the direction of positron beam.
Tracks (except for $\ks$ daughters) must also be consistent with coming from the interaction point (IP) in three dimensions.
Particle identification (PID) combining information of
measured energy loss (\dEdx) in the MDC and the flight time obtained
from the TOF is used to separate charged kaons and pions, the likelihood is required to be
$\mathcal{L}(K) > \mathcal{L}(\pi)$, $\mathcal{L}(K) > 0$ for kaons and vice-versa for pions.
Electromagnetic showers are reconstructed
by clustering EMC crystal energies; efficiency and
energy resolution are improved by including the energy deposited
in nearby TOF counters. To identify
photon candidates, showers must have minimum energies of 25 MeV
for $|\cos\theta|<0.80$ (barrel region) or 50 MeV for
$0.86<|\cos\theta|<0.92$ (endcap regions). The angle between the
shower direction and all track extrapolations to the EMC must be larger
than 10 standard deviations. A requirement on the EMC timing
suppresses electronic noise and energy deposits unrelated to the event.
The $\pi^0$ candidates are reconstructed by requiring the diphoton invariant
mass to obey $M_{\gamma \gamma}\in (0.115, 0.150) $  \GeV. Candidates with
both photons coming from the endcap regions are rejected due to poor resolution.
To improve resolution and reduce background, we constrain the invariant
mass of each photon pair to the nominal $\pi^0$ mass~\cite{pdg2012}.
The $\ks$  candidates are selected from pairs of
oppositely charged and vertex-constrained tracks consistent with coming from the IP
along the beam direction but free of aforementioned PID and having an invariant mass
in the range  $0.487 < M_{\pip\pim} < 0.511$ \GeV.


The ST candidate events are selected by reconstructing a $D^-$ or $\bar{D}^0$  in the following hadronic final states:
$D^- \to K^+\pi^-\pi^-$, $K^+\pi^-\pi^-\pi^{0}$, $\ks\pi^-$,
$\ks\pi^-\pi^{0}$, $\ks\pi^+\pi^-\pi^-$, $K^+K^-\pi^-$,
and $\bar{D}^0 \to K^+\pi^-$, $K^+\pi^-\pi^{0}$, $K^+\pi^-\pi^+\pi^-$, $K^+\pi^-\piz\piz$, $K^+\pi^-\pi^+\pi^-\piz$,
comprising approximately 28.0\% and 38.0\%~\cite{pdg2012}
of all $D^-$ and $\bar{D}^0$ decays, respectively.
For the signal side, we reconstruct $D^+ \to  \omega\pi^+ (\eta \pi^+)$  and $D^0 \to \omega \piz (\eta \piz)$, with $\omega (\eta) \to \pi^+\pi^- \pi^0$.
Throughout the paper, charge-conjugate modes are implicitly implied, unless otherwise noted.

To identify the reconstructed $D$ candidates, we use two variables,
the beam-constrained mass, $\mbc$, and the energy difference, $\delE$,
which are defined as
\begin{equation}
\mbc \equiv \sqrt{E_{\text{beam}}^{2}/c^{4}-|\vec{p}_{D}|^{2}/c^{2}}, \,\, \Delta E\equiv E_{D}-E_{\rm beam} \,.
\end{equation}
Here, $\vec{p}_{{D}}$ and ${E}_{D}$ are the reconstructed  momentum and energy of the
$D$ candidate in the $e^+e^-$ center-of-mass system, and $E_{\rm beam}$
is the beam energy.  For true $D^{+,0}$ candidates, $\Delta E$  will be
consistent with zero, and $M_{\text{BC}}$ consistent with the $D^{+,0}$ mass.
The resolution of $M_{\rm BC}$ is less than 2 MeV/$c^2$ and is dominated by
the beam energy spread.  The $\Delta E$ resolution is about $10$~MeV
for final states consisting entirely of charged tracks, but increases to
about $15$ ($20$)~MeV for cases where one (two) $\pi^0$ are included.
We accept $D$ candidates with $M_{\rm BC}$ greater than 1.83 GeV/$c^2$ and with mode-dependent $\Delta E$ requirements
of approximately three standard deviations ($\sigma$) around the fitted double Gaussian means.
For the ST modes, we accept at most one candidate per mode per event;
the candidate with the smallest $|\Delta E |$ is chosen~\cite{He:2005bs}.

\begin{figure}
\begin{center}
\includegraphics[height=7cm]{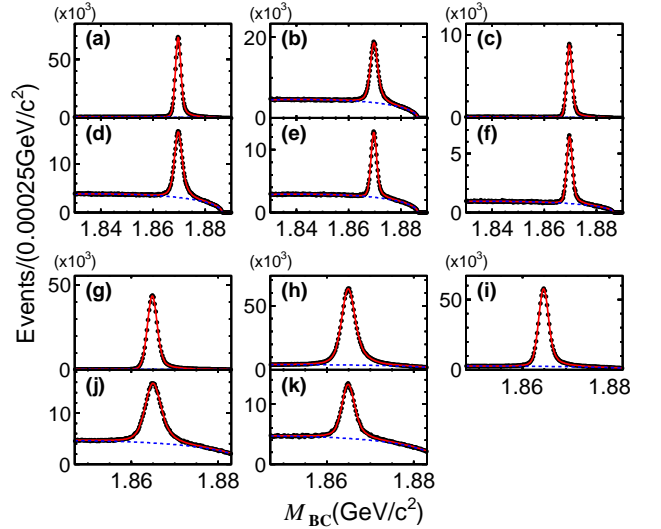}
\caption{$\mbc$ distributions of ST samples for different tag modes.
The first two rows show charged\dm decays: (a) $K^+\pi^-\pi^-$, (b) $K^+\pi^-\pi^-\pi^{0}$, (c) $\ks\pi^-$,
(d) $\ks\pi^-\pi^{0}$, (e) $\ks\pi^+\pi^-\pi^-$, (f) $K^+K^-\pi^-$,
the latter two rows show neutral\dm decays: (g) $K^+\pi^-$, (h) $K^+\pi^-\pi^{0}$, (i) $K^+\pi^-\pi^+\pi^-$, (j) $K^+\pi^-\piz\piz$, (k) $K^+\pi^-\pi^+\pi^-\piz$.
Data are shown as points, the (red) solid lines are the total fits
and the (blue) dashed lines are the
background shapes. $D$ and $\bar{D}$ candidates are combined.}
\label{fig:singleTagSample}
\end{center}
\end{figure}

To obtain ST yields, we fit the $M_{\rm BC}$ distributions of the accepted $D$ candidates,
as shown in Fig.~\ref{fig:singleTagSample}.
The signal shape which is modeled by MC shape convoluted with a Gaussian function
includes the effects of beam energy spread, ISR, the $\psi(3770)$  line shape, and resolution.
Combinatorial background is modeled by an ARGUS function~\cite{Albrecht:1990am}.
With requirement of  $1.866<M^{\rm tag}_{\rm BC}<1.874$~\GeV~for $D^+$ case or  $1.859<M^{\rm tag}_{\rm BC}<1.871$~\GeV~for $D^0$ case, ST yields are calculated by subtracting the integrated ARGUS background yields within the signal
region from the total event counts in this region. The tag efficiency is
studied using MC samples following the same procedure.
The ST yields in data and corresponding tag efficiencies are listed
in Table~\ref{tab:tagEfficiency}.

On the signal side we search for $D^+ \to \pi^+\pi^- \pi^0 \pi^+$ and $D^0 \to \pi^+ \pi^-\pi^0\pi^0$ modes containing an $\omega(\eta) \to \pi^+\pi^-\pi^0$ decay.
The requirements on $\Delta E$
are applied similar as in the tag selection; if multiple candidates
are found, the candidate with the minimum $|\Delta E|$ is chosen.
For both $D^+$ and $D^0$ decays, two possible $\omega$ $(\eta)$ combinations
exist.  Combinations with $3\pi$ mass in the interval $(0.4,1.0)$~\GeV~
are considered. The chance that both $\omega$ $(\eta)$ candidates combinations
lie in this region is only about $0.3\%$, rendering this source of multiple
candidates negligible.

With the DT technique, the continuum background $e^+e^- \to q\bar{q}$ is highly suppressed. The remaining background dominantly comes from $D\bar{D}$ events
broadly populating the $3\pi$ mass window.  To suppress the non-$\omega$ background, we require that the helicity, $H_{\omega} \equiv \rm cos \theta_{H}$, of the
$\omega$ have an absolute value larger than 0.54~(0.51) for $D^+$ ($D^0$).
The angle $\theta_{H}$ is the opening angle between the direction of the normal
to the $\omega\to 3\pi$ decay plane and direction of the $D$ meson
in the $\omega$ rest frame.
True $\omega$ signal from $D$ decays is longitudinally polarized so we expect a $\rm{cos^{2} \theta_{H}} \equiv$ $H^{2}_{\omega}$ distribution.
To further suppress  background from $D^{+,0}\to \ks\pi^+\pi^{0,-}$ with $\ks \to \pi^+\pi^-$, we apply a $\ks$ veto by requiring
$|M_{\pip\pim}-m^{\rm PDG}_{K^0_S}|>12~(9)$~\MeV~ for the $D^+$ $(D^0)$ analysis.
Here, $m^{\rm PDG}_{K^0_S}$ is the known $\ks$ mass and $M_{\pi^+\pi^-}$ is calculated at the IP for simplicity. The requirements on the $\omega$ helicity and $\ks$ veto are optimized to get maximum sensitivity based on the signal MC events and data in $\omega$ sidebands.

\begin{figure}[hbtp]
\begin{center}
\includegraphics[height=4cm]{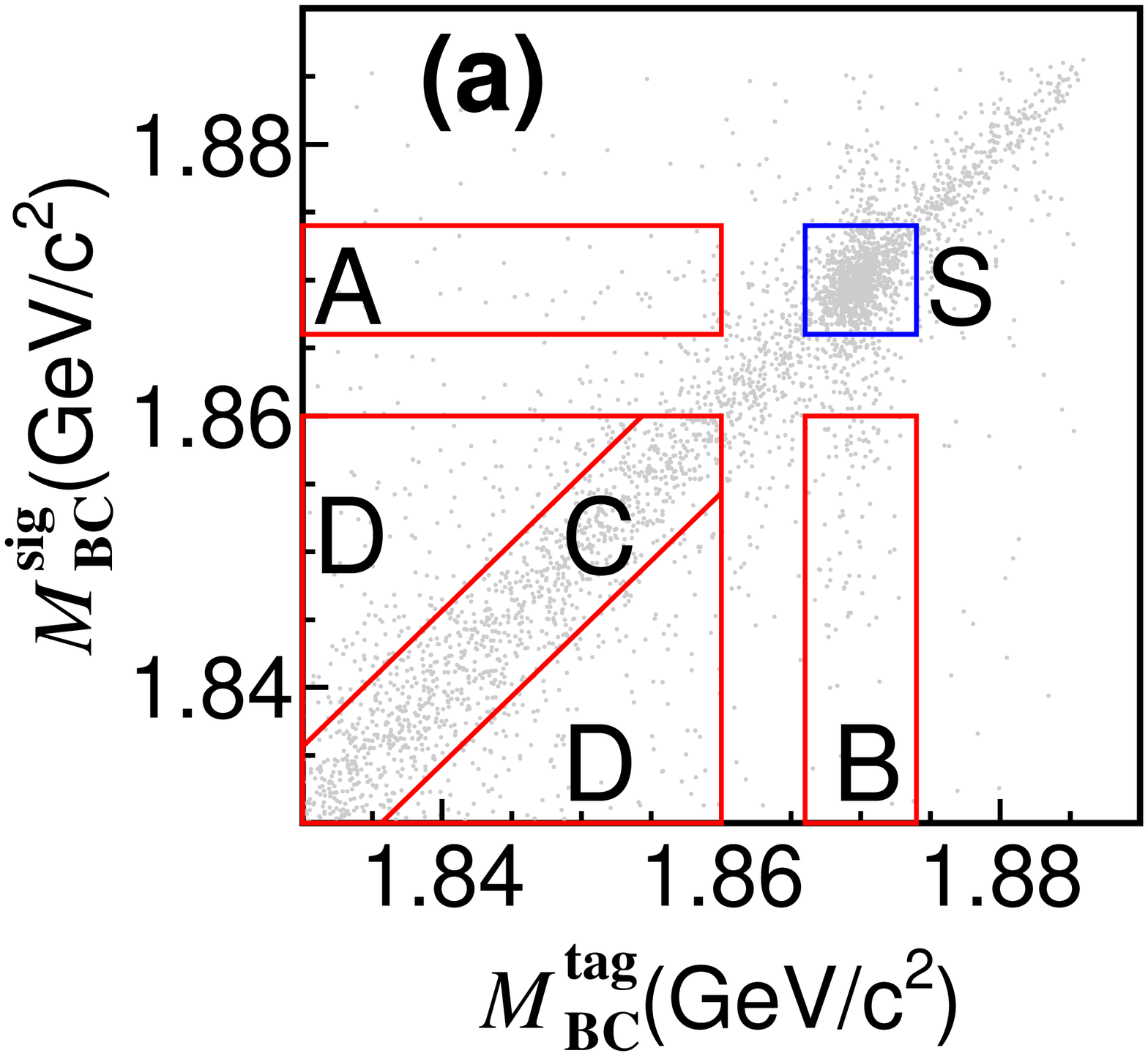}
\includegraphics[height=4cm]{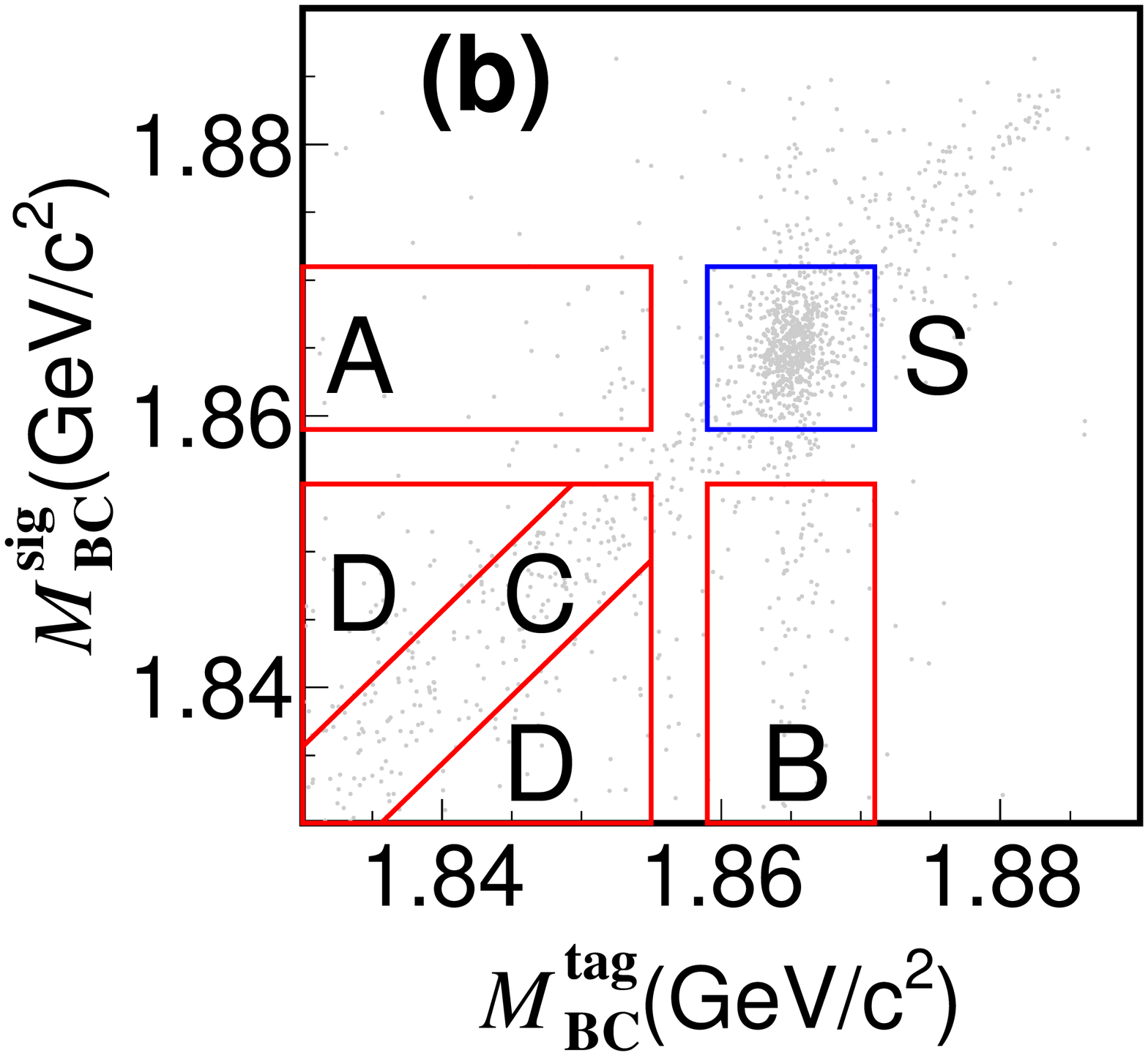}
\caption{2D $M_{\rm BC}$ distributions for
(a) $D^+ \to \omega \pi^+$  and (b) $D^0 \to \omega \pi^0$
with the signal ({\bf S}) and sideband ({\bf A, B, C, D}) regions
used for background estimation indicated.}
\label{fig:boxDefinition}
\end{center}
\end{figure}

After the above selection criteria, the signal region {\bf S}  for the DT candidates is defined as $1.866 < M_{\rm BC}< 1.874$~\GeV~for the $D^+$  ($1.859 < M_{\rm BC}< 1.871$~\GeV~for the $D^0$) in the two-dimensional (2D) $M^{\rm sig}_{\rm BC}$ versus $M^{\rm tag}_{\rm BC}$ plane, as illustrated in Fig.~\ref{fig:boxDefinition}.
We also define sideband box regions to estimate potential background~\cite{ref2DRegion}.
Sidebands {\bf A} and {\bf B} contain candidates  where either the $D$ or the $\bar{D}$ is misreconstructed.
Sidebands {\bf C} and {\bf D} contain candidates where both $D$ and $\bar{D}$  are misreconstructed, either in a correlated way ({\bf C}), by assigning daughter particles to the wrong parent, or in an uncorrelated way ({\bf D}).

\begin{figure}
\begin{center}
\includegraphics[height=4.1cm]{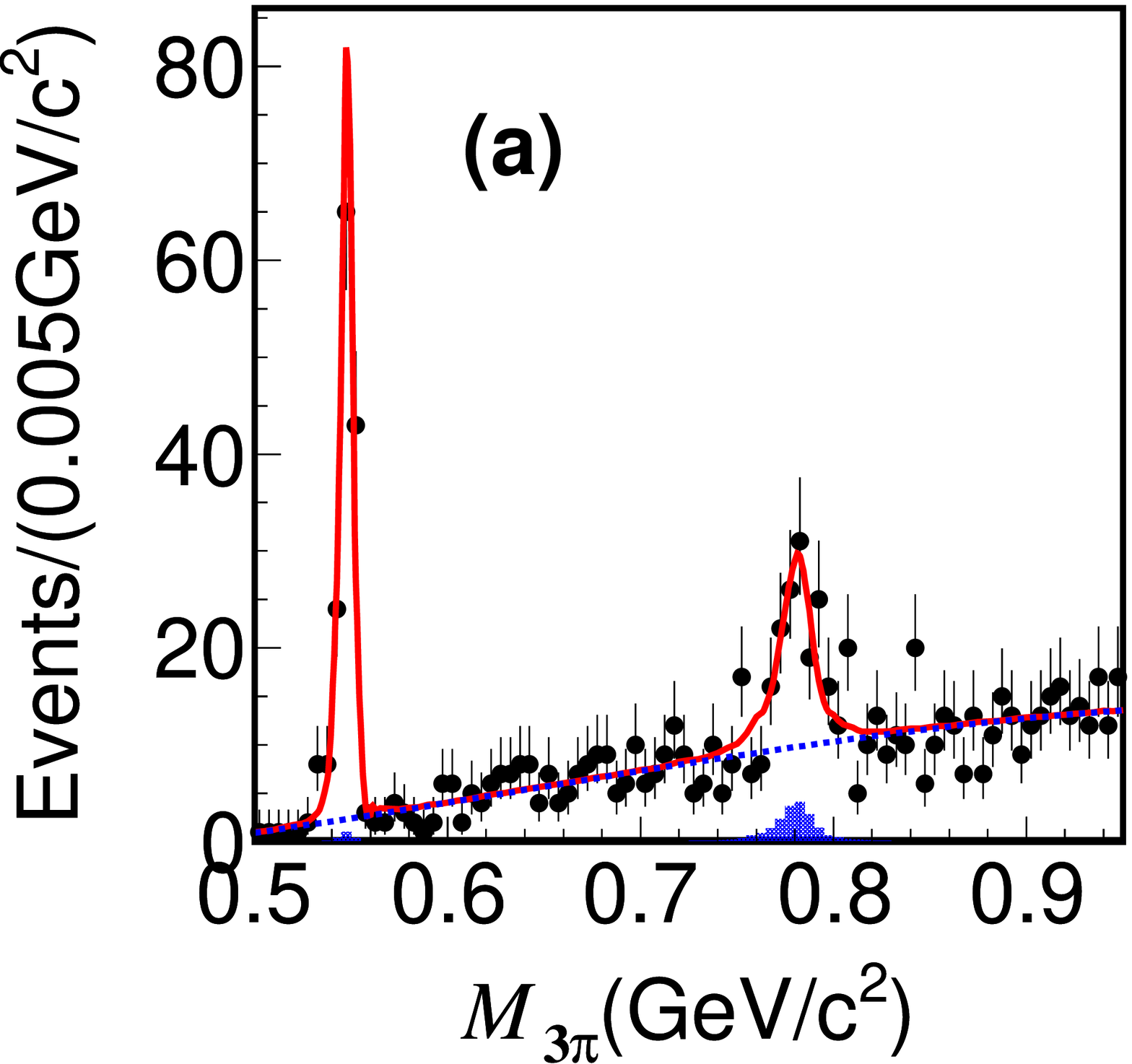}
\includegraphics[height=4.1cm]{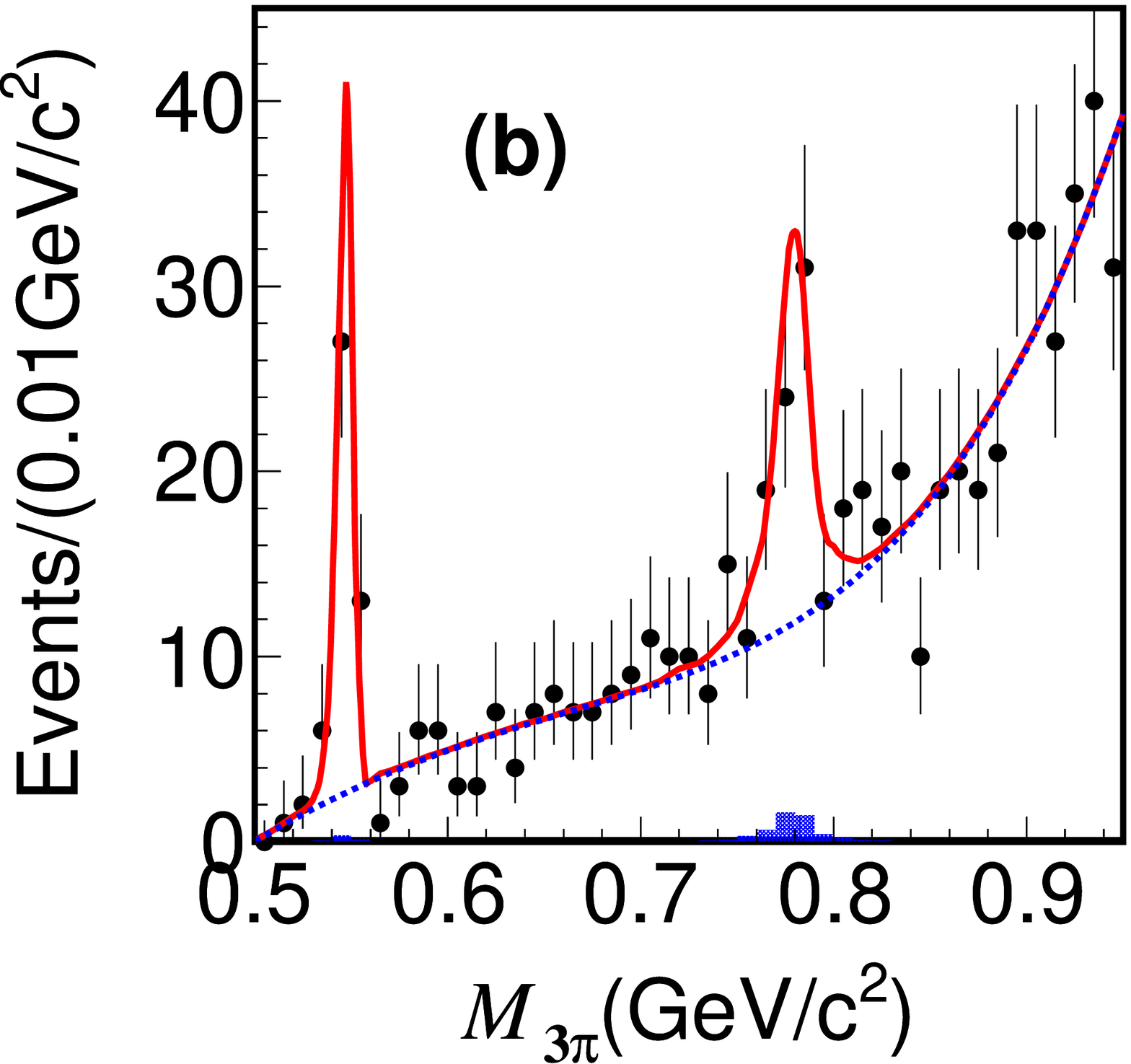}
\caption{Fits to the $3\pi$ mass spectra for (a) $D^+\to \pi^+\pi^-\pi^0\pi^+$
and (b) $D^0 \to \pi^+\pi^-\pi^0\pi^0$ in the signal region {\bf S} as
defined in Fig.~\ref{fig:boxDefinition}.  Points are data; the (red)
solid lines are the total fits; the (blue) dashed lines
are the background shapes, and the hatched histograms
are peaking background estimated from 2D $M_{\rm BC}$ sidebands.}
\label{fig:nominalFit}
\end{center}
\end{figure}

\begin{table}[hbtp]
\caption{ST data yields ($N_{\rm tag}^{\rm obs}$), ST ($\epsilon_{\rm tag}$) and DT
($\epsilon^\omega_{\rm tag, sig}$ and $\epsilon^\eta_{\rm tag, sig}$)
efficiencies, and their statistical uncertainties.
Branching fractions of the $\ks$ and $\piz$  are not included in the efficiencies, but are included in the branching fraction calculations. The first six rows are for $D^-$ and the last five are for $\bar{D}^0$. }
\label{tab:tagEfficiency}
\begin{center}
\begin{footnotesize}
\renewcommand{\arraystretch}{1.0}
\scalebox{0.92}{
\begin{tabular}{ccccc}
\hline \hline
Mode                 &  ST Yields       & $\epsilon_{\rm tag} $ $(\%)$   &  $\epsilon^\omega_{\rm tag, sig}$$(\%)$ & $\epsilon^\eta_{\rm tag, sig}$$(\%)$ \\
\hline
$K^+\pi^-\pi^-            $  &  $772711  \pm 895  $  &  $48.76  \pm 0.02 $ & $11.01 \pm 0.15 $ & $12.64 \pm 0.17 $ \\
$K^+\pi^-\pi^-\pi^{0}     $  &  $226969  \pm 608  $  &  $23.19  \pm 0.02 $ & $4.47  \pm 0.10 $ & $5.26  \pm 0.11 $ \\
$\ks\pim                  $  &  $95974   \pm 315  $  &  $52.35  \pm 0.07 $ & $11.69 \pm 0.18 $ & $13.99 \pm 0.21 $ \\
$\ks\pim\pi^{0}           $  &  $211872  \pm 572  $  &  $26.68  \pm 0.03 $ & $5.35  \pm 0.13 $ & $6.44  \pm 0.14 $ \\
$\ks\pim\pip\pim          $  &  $121801  \pm 459  $  &  $30.53  \pm 0.04 $ & $6.16  \pm 0.13 $ & $7.17  \pm 0.15 $ \\
$K^+K^-\pim               $  &  $65955   \pm 306  $  &  $38.72  \pm 0.07 $ & $8.50  \pm 0.13 $ & $9.76  \pm 0.14 $ \\
\hline
$K^+\pim                  $  &  $529558  \pm 745  $  &  $64.79  \pm 0.03 $ & $12.44 \pm 0.16 $ & $14.17 \pm 0.17 $ \\
$K^+\pim\pi^{0}           $  &  $1044963 \pm 1164 $  &  $34.13  \pm 0.01 $ & $5.73  \pm 0.11 $ & $6.87  \pm 0.12 $ \\
$K^+\pim\pip\pim          $  &  $708523  \pm 946  $  &  $38.33  \pm 0.02 $ & $6.04  \pm 0.11 $ & $7.00  \pm 0.13 $ \\
$K^+\pim\piz\piz          $  &  $236719  \pm 747  $  &  $13.87  \pm 0.02 $ & $1.78  \pm 0.06 $ & $2.10  \pm 0.07 $ \\
$K^+\pim\pip\pim\piz      $  &  $152025  \pm 684  $  &  $15.55  \pm 0.03 $ & $1.93  \pm 0.06 $ & $2.08  \pm 0.07 $ \\
\hline
\hline
\end{tabular}
}
\end{footnotesize}
\end{center}
\end{table}
%
To obtain the $\omega (\eta)$ yield, we perform a fit to the  $\pip\pim\piz$ invariant mass
$(M_{3\pi})$ distribution with events in the signal region {\bf S}. The $\omega(\eta)$
shape is modeled by the signal MC shape convoluted with
a  Gaussian function to describe the difference in the $M_{3\pi}$
resolution between MC and data.  Due to high statistics, the width
$\sigma_\eta$ of the Gaussian for the $\eta$ case is determined by the fit, while the width $\sigma_\omega$ for the $\omega$ case
is constrained by the MC-determined ratio $R = \sigma^{\rm MC}_\omega/\sigma^{\rm MC}_\eta$ giving the relative $M_{3\pi}$ resolution for $\eta$ and $\omega$
final states.
For $D^+$, the background shape is described by a third-order Chebychev polynomial, while for $D^0$ we use a shape
of $a_{0}M_{3\pi}^{1/2}+a_{1}M_{3\pi}^{3/2}+a_{2}M_{3\pi}^{5/2}+a_{3}M_{3\pi}^{7/2}+a_{4}M_{3\pi}^{9/2}$
, where $a_i$ ($i= 0,\ldots,4$) are free parameters.  The fit
results are shown in Fig.~\ref{fig:nominalFit}, and the total $\omega
$ yields $N_{\omega}$ for $D^+$ and $D^0$ cases are listed in
Table~\ref{tab:yield}.

To estimate the $\omega(\eta)$ yield in the signal region {\bf S} from
background processes,
event counts in sidebands {\bf A}, {\bf B}, and {\bf C} are projected into the signal region {\bf S} using scale factors determined from integrating the background shape in the ST $M_{\rm BC}$ fits.
Contributions to sideband {\bf D} are assumed to be uniformly distributed across the other regions~\cite{ref2DRegion}.  For these events from the sideband regions, we perform similar fits to the $3\pi$ mass spectra, and find the
peaking background yields $N^{\rm bkg}_{\omega(\eta)}$ for $D^+$ and $D^0$ respectively, as listed in Table~\ref{tab:yield}.
By subtracting the $\omega$ peaking background extending underneath the signal region, the DT signal yields, $N^{\rm obs}_{\rm sig}$, are obtained. The statistical significances for $\dpomegapi$ and $\dzomegapi$ are found to be
$\sigDpStat$ and $\sigDzStat$, respectively, as determined by the ratio of the nominal maximum likelihood value and the likelihood value for a fit where the signal is set to zero by fixing the total yield $N_\omega$ to be equal to the
sideband based background prediction, $N^{\rm bkg}_{\omega(\eta)}$.

\begin {table}[hbtp]
\caption{Summary for the total $\omega$ ($\eta$)  yields ($N_{\omega (\eta)}$), $\omega (\eta)$ peaking background yields ($N^{\rm bkg}_{\omega(\eta)}$)
and net DT yields ($N^{\rm obs}_{\rm sig}$) in the signal region {\bf S} as defined in
Fig. ~\ref{fig:boxDefinition}. $N^{\rm obs}_{\rm sig}$ is estimated from the defined sidebands. The errors are statistical.
 }
\label{tab:yield}
\begin {center}
\begin {small}
\renewcommand{\arraystretch}{1.4}
\begin {tabular}{cccc}
\hline \hline
Mode            &   $N_{\omega (\eta)}$    &  $N^{\rm bkg}_{\omega(\eta)}$  &  $N^{\rm obs}_{\rm sig}$\\
\hline
$D^+ \to \omega\pip$     &   $100\pm16$ & $21\pm4$    &  $79\pm16$ \\
$D^0 \to \omega\piz$     &   $50\pm12$ & $5\pm5$     &  $45\pm13$ \\
$D^+ \to \eta \pip$     &   $264\pm17$ & $6\pm2$   &  $258 \pm 18$ \\
$D^0 \to \eta\piz$     &   $78\pm10$  & $3\pm2$   &  $75 \pm 10$ \\

\hline
\hline
\end {tabular}
\end {small}
\end {center}
\end {table}

\begin{figure}
\begin{center}
\includegraphics[height=3cm]{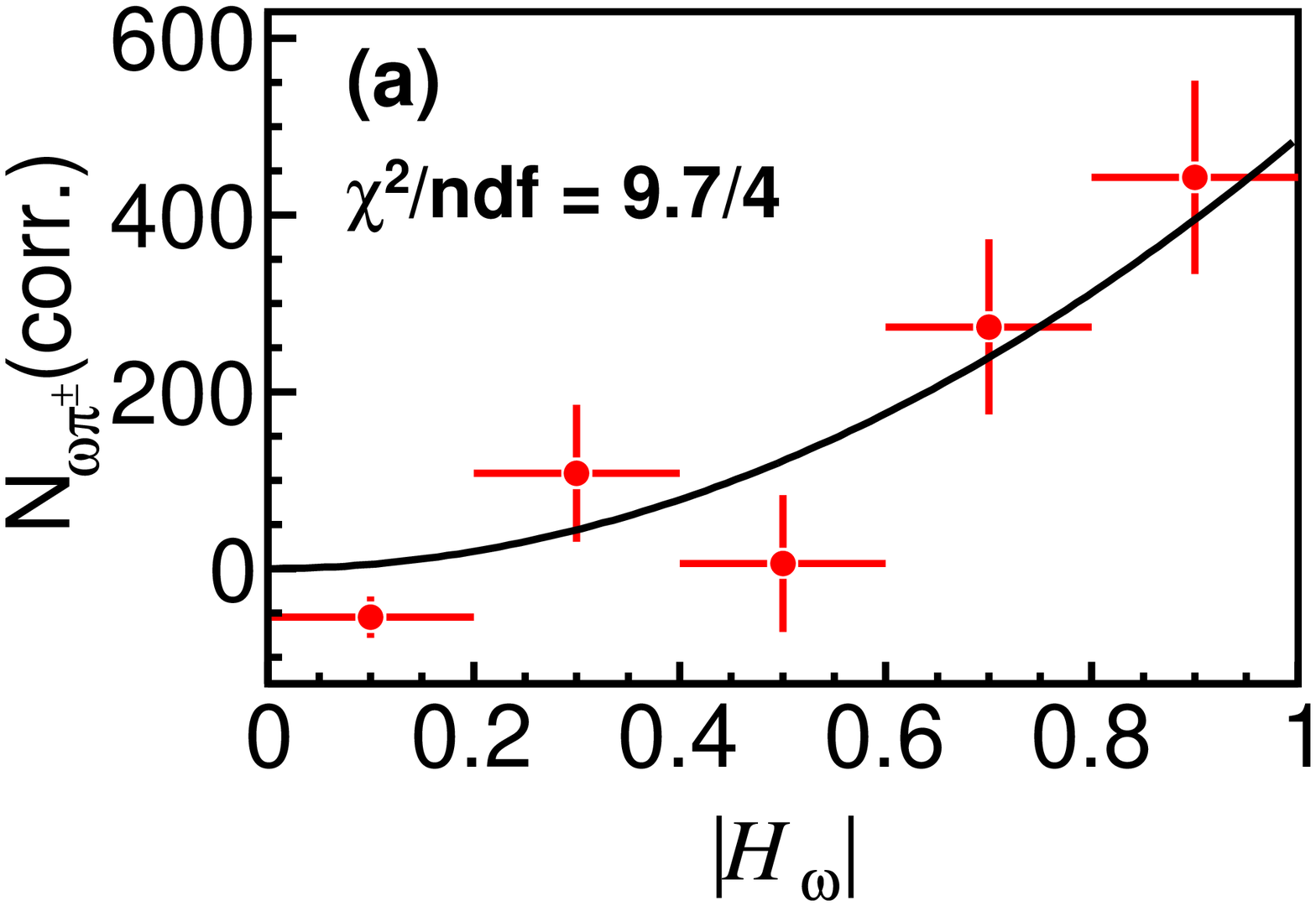}
\includegraphics[height=3cm]{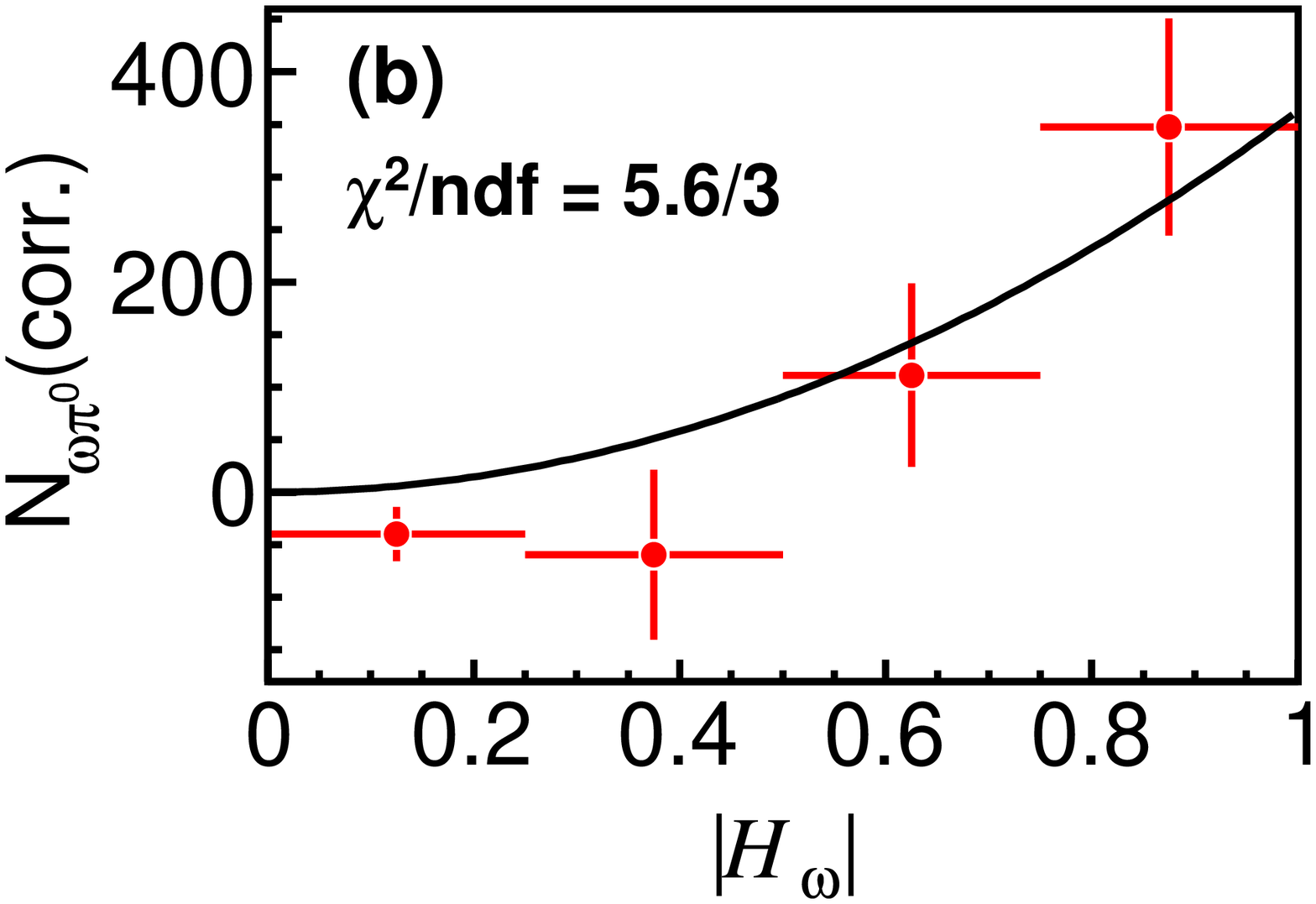}
\caption{ Efficiency corrected yields versus $|H_\omega|$ for (a) $D^+
  \to \omega \pi^+$  and (b) $D^0 \to \omega \pi^0$. Both are
  consistent with a distribution like $\cos^2{\theta_{H}}$ (black line).}
\label{fig:helicity_dep}
\end{center}
\end{figure}

We now remove the $\omega$ helicity requirement, and investigate the helicity
dependence of our signal yields. By following procedures similar to those
described above, we obtain the signal yield in each $|H_{\omega}|$ bin.
The efficiency corrected yields are shown in Fig.~\ref{fig:helicity_dep},
demonstrating agreement with expected $\cos^2 \theta_{H}$ behavior,
further validating this analysis.

With analogous selection criteria, we also determine $\BR(D^{+,0} \to \eta\pi^{+,0})$
as a cross-check.  The results are found to be consistent with the
nominal results given below for $\BR(D^{+,0} \to \eta\pi^{+,0})$,
using relaxed cuts, as well as the PDG listings~\cite{pdg2012}.

\begin{figure}
\begin{center}
\includegraphics[height=4cm]{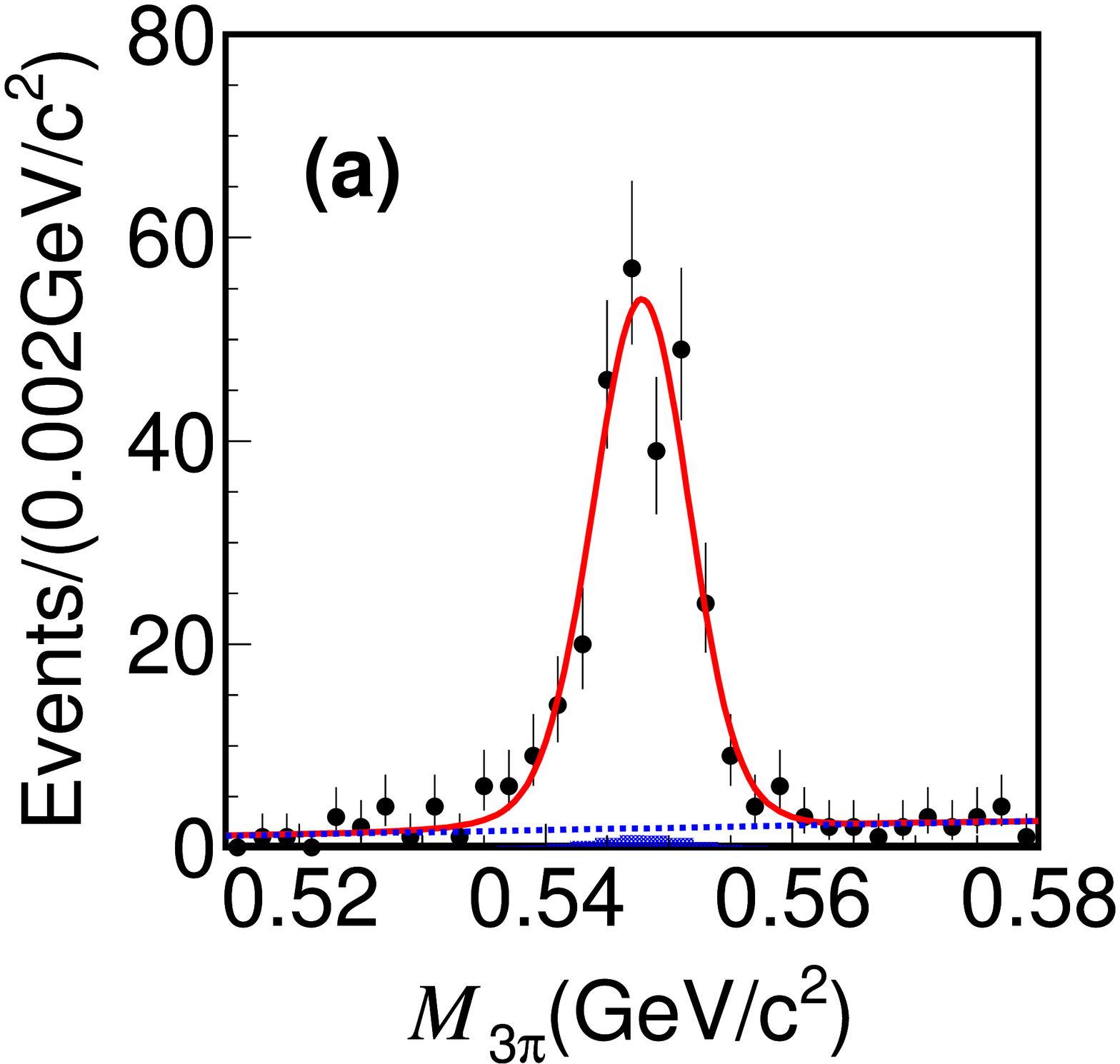}
\includegraphics[height=4cm]{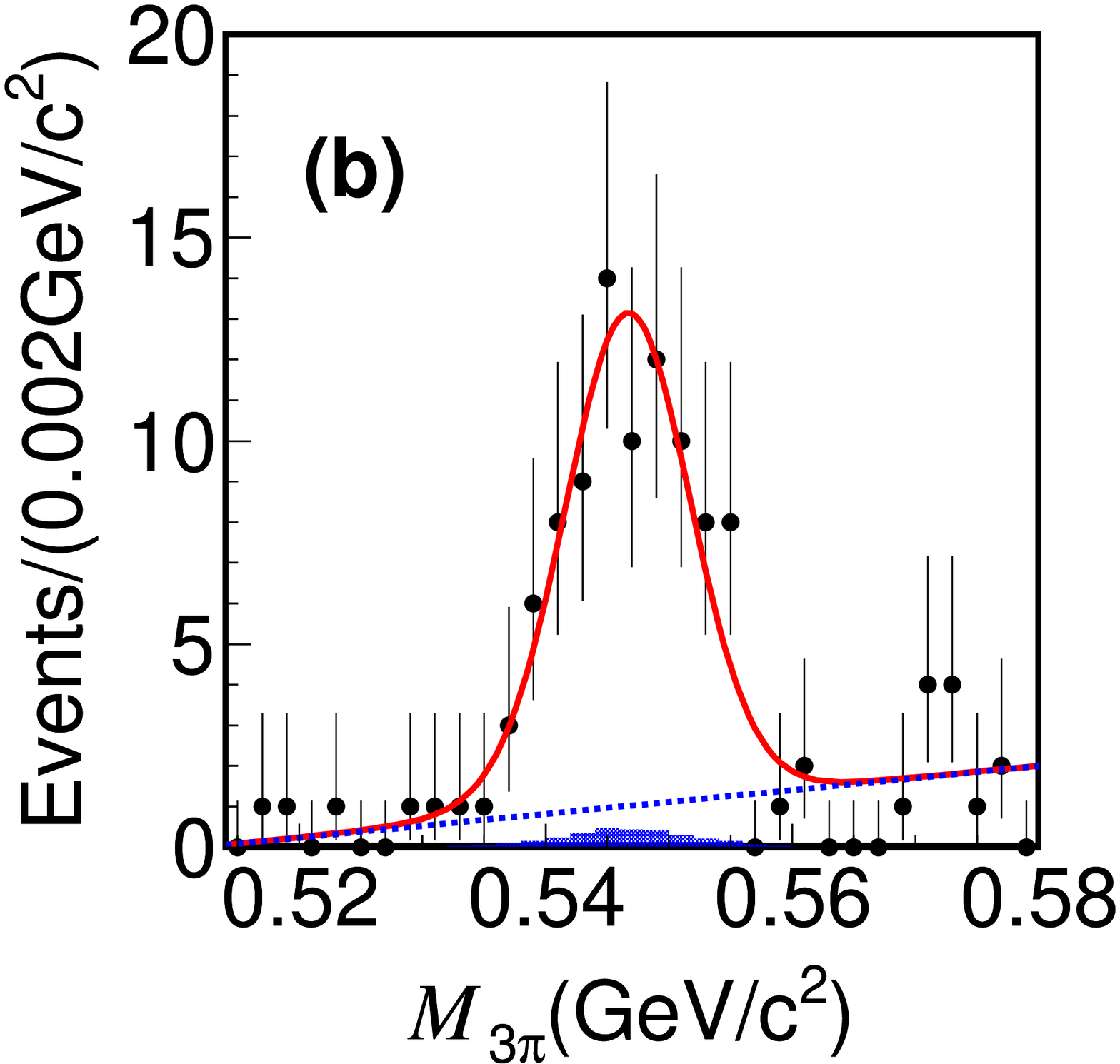}
\caption{Fits to the $3\pi$ mass spectra for (a) $D^+\to \pi^+\pi^-\pi^0\pi^+$
and (b) $D^0 \to \pi^+\pi^-\pi^0\pi^0$ in the $\eta$ mass region for the
signal region {\bf S} as defined in Fig.~\ref{fig:boxDefinition}.
Points are data; the (red) solid lines are the total fits;
the (blue) dashed lines are the background shapes, and the hatched histograms
are peaking background estimated from 2D $M_{\rm BC}$ sidebands.}
\label{fig:etaResult}
\end{center}
\end{figure}

As shown in Fig.~\ref{fig:nominalFit},  the background level in the $\eta$ signal region of the $3\pi$ invariant mass distribution is small compared to that near the $\omega$ mass.
Also, according to the MC simulations and fits to events from the 2D $M_{\rm BC}$ sideband regions, $\eta$ peaking background is small, as shown
in Fig.~\ref{fig:nominalFit}.  Therefore, to improve statistics, we remove the $\ks$ veto requirements and also make no helicity requirement since
$H_\eta\equiv \rm cos \theta_{H}$ for signal is flat.
Following a similar fit procedure, with results shown in Fig.~\ref{fig:etaResult}, we determine $\eta\pi^+$ and $\eta\piz$ DT yields as listed in Table~\ref{tab:yield}.

With the DT technique, the branching fraction measurements are insensitive to
systematics coming from the ST side since they mostly cancel.
For the signal side, systematic uncertainties mainly come from imperfect
knowledge of the efficiencies for tracking finding, PID criteria,
the $\ks$ veto, and the $H_{\omega}$ requirement;
additional uncertainties are related to the fit procedures.

Possible differences in  tracking, PID and
$\piz$ reconstruction efficiencies between data and the MC simulations
are investigated using a partial-reconstruction technique based on the control
samples $D^0 \ar K^-\pi^+\piz$ and $D^0 \ar K^-\pi^+$.  We assign
uncertainties of $1.0\%$ and $0.5\%$ per track for track finding and PID, respectively,
and 1.0\% per reconstructed $\piz$.

Uncertainty due to the 2D signal region definition is investigated via the
relative change in signal yields for different signal region definitions
based on the control samples $D^{+}\to \ks\pi^+\piz$ and
$D^{0}\to \ks\piz\piz$ which have the same pions in the final state
as our signal modes.
With the same control samples, uncertainties due to the $\Delta E$
requirements are also studied.  The relative data-MC efficiency differences
are taken as systematic uncertainties,
as listed in Table~\ref{tab:SummarySys}.

Uncertainty due to the $|H_{\omega}|$ requirement is studied using the
control sample $D^{0}\to \ks\omega$. The data-MC efficiency difference
with or without this requirement is taken as our systematic.
Uncertainty due to the $\ks$ veto is similarly obtained with
this control sample.

The $\omega$ peaking background is estimated from 2D $M_{\rm BC}$ sidebands. We change the sideband ranges
by 2~MeV/$c^{2}$ for both sides and investigate the fluctuation on the  signal yields,
which is taken as a systematic uncertainty.

In the nominal fit to the $M_{3\pi}$ distribution,  the ratio $R$, which is the relative difference on the $M_{3\pi}$ resolution between $\eta$ and $\omega$ positions, is determined by MC simulations.
With control samples  $D^0 \to \ks\eta$
and $\ks\omega$, the difference between data and MC defined as
$\delta R=R_{\rm data}/R_{\rm MC}-1$ is obtained.
We vary the nominal $R$ value by $\pm1\sigma$ and take the relative change
of signal yields as a systematic uncertainty.

Uncertainties due to the background shapes are investigated by
changing the orders of the polynomials employed.
Uncertainties due to the $M_{3\pi}$ fitting range are investigated by changing the range
from $(0.50, 0.95)$~\GeV~ to $(0.48, 0.97)$~\GeV~in the fits, yielding
relative differences which are taken as systematic uncertainties.

We summarize the systematic uncertainties in Table~\ref{tab:SummarySys}.
The total effect is calculated by combining the uncertainties
from all sources in quadrature.

\begin {table}[hbtp]
\caption{Summary of systematic uncertainties in \%. Uncertainties
which are not involved are denoted by ``--''.}
\label{tab:SummarySys}
\begin {center}
\begin {footnotesize}
\begin {tabular}{ccccc}
\hline \hline
Source                                            &   $\omega\pi^{+}$        &   $\omega\piz$   & $\eta\pi^{+}$            & $\eta\piz$\\
\hline
 $\pi^\pm$ tracking                               &   3.0                    &  2.0             &   3.0                    &  2.0  \\
$\pi^\pm$ PID                                     &   1.5                    &  1.0             &   1.5                    &  1.0  \\
$\piz$ reconstruction                             &   1.0                    &  2.0             &   1.0                    &  2.0  \\

2D $\mbc$ window                                  &   0.1                    &  0.2             &   0.1                    &  0.2  \\
$\Delta E$ requirement                            &   0.5                    &  1.6             &   0.5                    &  1.6  \\
$|H_{\omega}|$ requirement                        &   3.4                    &  3.4             &   --                     &  --   \\
$\ks$ veto                                        &   0.8                    &  0.8             &   --                     &  --   \\

Sideband regions                                  &   1.3                    &  2.2             &   0.0                    &  0.5  \\

Signal resolution                                 &   0.9                    &  0.9             &  --                      &  --   \\
Background shape                                  &   2.3                    &  1.3             &  1.9                     &  3.5  \\
Fit range                                         &   0.3                    &  1.9             &  0.8                     &  1.5  \\

$\BR(\omega(\eta)\ar\pip\pim\piz)$~\cite{pdg2012}  &   0.8                    &  0.8             &  1.2                     & 1.2  \\
\hline
Overall                                           &   5.8                    &  6.0             &  4.3                     & 5.3  \\
\hline\hline
\end {tabular}
\end {footnotesize}
\end {center}
\end {table}

Finally, the measured branching fractions of $D \to \omega\pi$ and $\eta \pi$ are summarized in Table~\ref{tab:SummaryBrs}, where the first errors are statistical and the second ones are systematic.

\begin {table}[hbtp]
\caption{Summary of branching fraction measurements, and comparison with the previous measurements for $D\to \omega \pi$~\cite{4piCLEO} and $D\to \eta \pi$~\cite{etapiCLEOc}. }
\label{tab:SummaryBrs}
\begin {center}
\begin {footnotesize}
\begin {tabular}{ccccc}
\hline \hline
Mode                                &   This work       &   Previous measurements\\
\hline
$\dpomegapi$                        &   $(\resultBrDp)\times 10^{-4}$       &  $<3.4\times10^{-4}$ at $90\%$ C.L. \\
$\dzomegapi$                        &   $(\resultBrDz)\times 10^{-4}$       &  $<2.6\times10^{-4}$ at $90\%$ C.L. \\
$\dpetapi$                          &   $(\resultBrDpEta)\times10^{-3}$     &  $(\BrDpEtaPDG)\times10^{-3}$  \\
$\dzetapi$                          &   $(\resultBrDzEta)\times10^{-3}$     &  $(\BrDzEtaPDG)\times10^{-3}$  \\
\hline\hline
\end {tabular}
\end {footnotesize}
\end {center}
\end {table}

In summary, we present the first observation of the SCS
decay $\dpomegapi$ with statistical significance of $\sigDpStat$. We find the first evidence for the SCS decay $\dzomegapi$ with statistical significance of
 $\sigDzStat$. The results are consistent with the theoretical prediction~\cite{HYCheng},
and can improve understanding of $U$-spin and $SU(3)$-flavor symmetry
breaking effects in $D$ decays~\cite{Grossman:2012ry}.
We also present measurements of the branching fractions for $\dpetapi$
and $\dzetapi$ which are consistent with the previous measurements~\cite{etapiCLEOc}.

\vspace{0.9cm}

The BESIII collaboration thanks the staff of BEPCII and the IHEP computing center for their strong support. This work is supported in part by National Key Basic Research Program of China under Contract No. 2015CB856700; National Natural Science Foundation of China (NSFC) under Contracts Nos. 11125525, 11235011, 11322544, 11335008, 11425524; the Chinese Academy of Sciences (CAS) Large-Scale Scientific Facility Program; the CAS Center for Excellence in Particle Physics (CCEPP); the Collaborative Innovation Center for Particles and Interactions (CICPI); Joint Large-Scale Scientific Facility Funds of the NSFC and CAS under Contracts Nos. 11179007, 10975093, U1232201, U1332201; CAS under Contracts Nos. KJCX2-YW-N29, KJCX2-YW-N45; 100 Talents Program of CAS; National 1000 Talents Program of China; INPAC and Shanghai Key Laboratory for Particle Physics and Cosmology; German Research Foundation DFG under Contract No. Collaborative Research Center CRC-1044; Istituto Nazionale di Fisica Nucleare, Italy; Joint Funds of the National Science Foundation of China under Contract No. U1232107; Ministry of Development of Turkey under Contract No. DPT2006K-120470; Russian Foundation for Basic Research under Contract No. 14-07-91152; The Swedish Resarch Council; U. S. Department of Energy under Contracts Nos. DE-FG02-04ER41291, DE-FG02-05ER41374, DE-SC0012069, DESC0010118; U.S. National Science Foundation; University of Groningen (RuG) and the Helmholtzzentrum fuer Schwerionenforschung GmbH (GSI), Darmstadt; WCU Program of National Research Foundation of Korea under Contract No. R32-2008-000-10155-0.

\end{document}